\documentclass[twocolumn,twoside,slac_two]{revtex4}
\usepackage{graphicx}
\usepackage{fancyhdr}
\newcommand{\aap}{A\&A}
\newcommand{\apj}{ApJ}
\newcommand{\mnras}{MNRAS}

\newcommand{\nat}{Nature}
\newcommand{\beq}{\begin{equation}}
\newcommand{\eeq}{\end{equation}}
\newcommand{\beqn}{\begin{eqnarray}}
\newcommand{\eeqn}{\end{eqnarray}}
\newcommand{\lppr}{\stackrel{<}{\scriptstyle \sim}}

\newcommand{\pd}{\partial}

\begin{document}
\title{Particle acceleration timescales in relativistic shear flows}
\author{Frank M. Rieger and Peter Duffy}
\affiliation{Department of Mathematical Physics, University College Dublin,
 Belfield, Dublin 4, Ireland.}
\begin{abstract}
We review the acceleration of energetic particles in relativistic astrophysical 
jets characterized by a significant velocity shear. The possible formation of
power-law momentum spectra is discussed and typical acceleration timescales
are determined for a variety of different conditions such as parallel and 
azimuthal shear flows. Special attendance is given to the analysis of parallel 
shear flows with either a linear decreasing or a Gaussian-type velocity profile. 
It is shown that in the presence of a gradual shear flow and a particle mean free 
path scaling with the gyroradius, synchrotron radiation losses may no longer be 
able to stop the acceleration once it has started to work efficiently. 
Finally, the relevance of shear acceleration in small- and large-scale 
relativistic jets is addressed.
\end{abstract}

\maketitle

\thispagestyle{fancy}

\section{Introduction}
Relativistic jet outflows are observed across a wide range of astrophysical 
scales, from galactic microquasars to GRBs and radio-loud AGNs \cite{bri84,mir99,
zen97}. The fact that velocity gradients are a generic feature of environments 
producing jets suggests that many (if not all) of these outflows may possess a  
significant internal velocity shear. Today, the phenomenological evidence for 
internal jet stratification is indeed mounting and comprises evidence for universal 
structured jets in GRBs \cite{ros02,kum03}, internal jet rotation in AGNs \cite{rie02,
rie04}, and a multi-component jet structure consisting (at least) of a fast moving 
inner spine and a slower shear layer as indicated, for example, by the detailed 
analysis of the brightness and polarization systematics in kpc FR~I jets \cite{lai02,
lai04} or the intensity and polarization maps of pc-scale FR~I + II jets \cite{swa98,
per99,edw00}. It seems likely that in the presence of such a velocity shear efficient 
acceleration of particles to high energy may occur, provided particles are scattered 
across the shear by magnetic inhomogeneities carried within the flow \cite{rie04,
rie05}. The theory of shear acceleration deals with those cases where the impact of 
the systematic velocity components of the scattering centers becomes dominant over 
their random motion components known as the source of second-order Fermi acceleration 
effects.\\
Substantial theoretical contributions in the field of shear acceleration have been 
given by several authors (cf. \cite{rie04} for a review): Berezhko \& Krymskii~(1981), 
for example, showed that (non-relativistic gradual) shear acceleration can lead to 
power law particle momentum distributions resembling those of the classical Fermi 
theory, whereas Earl, Jokipii \& Morfill~(1988) derived the corresponding diffusive 
particle transport equation in the diffusion approximation including shear and 
inertial effects. The generalisation of the work by Earl et al. to the relativistic 
regime was achieved by Webb \cite{web85,web89,web94} using a mixed-frame 
approach and applied by Rieger \& Mannheim~(2002) to rotating and shearing jets in 
AGNs. Particle acceleration in non-gradual relativistic shear flows (i.e., in the 
presence of a relativistic velocity jump), on the other hand, has been analysed by 
Ostrowski \cite{ost90,ost98,ost00} using Monte Carlo methods, showing that very 
flat momentum spectra may occur.

\section{Microscopic picture}
Shear acceleration draws on a simple mechanism which seems to be applicable to 
a wide range of astrophysical flows such as accretion flows and structured jets 
in GRBs and AGNs (cf. also \cite{kat91}). According to the underlying physical
picture particles gain energy by scattering off (small-scale) magnetic field 
irregularities with different local velocities due to being systematically 
embedded in a collisionless velocity shear flow. The scattering process is 
assumed to occur in such a way that the particles are randomized in direction, 
with their energies being conserved in the local commoving fluid frame. In the 
presence of a velocity shear, the momentum of a particle travelling across the 
shear will change with respect to the local scattering frame so that for an 
isotropic particle distribution a net increase may occur \cite{jok90}. 
For illustration consider a continuous, non-relativistic shear flow with velocity
field given by
\beq
   \vec{u} = u_z(x)\,\vec{e}_z\,.
\eeq
Let $\vec{v}=(v_x,v_y,v_z)$ be the velocity vector, $m$ the relativistic mass and
$\vec{p}_1$ the initial momentum (relative to local flow frame) of the particle.
Within one scattering time $\tau$ (initially assumed to be independent of momentum)
the particle travels a distance $\delta x = v_x\,\tau$ across the shear, so that
in the presence of a gradual shear the flow velocity will have changed by $\delta
\vec{u} = \delta u\,\vec{e_z}$, with $\delta u = (\partial u_z/\partial x)\,\delta
x$. Hence the particle's momentum relative to the flow becomes $\vec{p}_2 =
\vec{p}_1 + m\,\delta \vec{u}$ (Galilean transformation), i.e.
\beq\label{trafo}
    p_2^2 = p_1^2 + 2\,m\,\delta u\,\,p_{1,z} + m^2\,(\delta u)^2\;.
\eeq As the next scattering event preserves the magnitude of the particle momentum
(in the local scattering frame) the particle magnitude will have this value in the 
local flow frame. Using spherical coordinates and averaging over solid angle 
assuming an almost isotropic particle distribution, the resultant average rate of 
momentum change and the average rate of momentum dispersion scale with the square 
of the flow velocity gradient \cite{jok90,rie05}, i.e., 
\beqn
  \left<\frac{\Delta p}{\Delta t}\right> 
       & \propto &  p \left(\frac{\pd u_z}{\pd x}\right)^2\,\tau \;, \label{p_dot}\\
  \left<\frac{\Delta p^2}{\Delta t}\right>  
       & \propto & p^2 \left(\frac{\pd u_z}{\pd x}\right)^2\,\tau \;.
\eeqn Generalizing these results to a momentum-dependent scattering time obeying 
a power law of the form $\tau \propto p^{\alpha}$, we may write down a simple 
Fokker-Planck transport equation for the phase space particle distribution function 
$f(p)$, assuming a mono-energetic injection of particles with momentum $p_0$. 
Solving for the steady-state with $\alpha > 0$ one obtains (see \cite{rie05})
\beq
     f(p) \propto p^{-\,(3 + \alpha)}\,\, H(p-p_0)\,,
\eeq where $H(p)$ is the Heaviside step function. Hence for a mean scattering time
scaling with the gyro-radius (Bohm case), i.e. $\tau \propto p^{\alpha}$, $\alpha 
=1$, one obtains $f(p) \propto p^{-4}$ , i.e., a power law particle number density 
$n(p)\propto p^{-2}$ which translates into a synchrotron emissivity $j_{\nu} 
\propto \nu^{-1/2}$. For a Kolmogorov-type ($\alpha =1/3$) or Kraichnan-type 
($\alpha=1/2$) scaling, on the other hand, much flatter spectra may be obtained, 
i.e., the synchrotron emissivity becomes $j_{\nu} \propto \nu^{-1/6}$ and 
$j_{\nu} \propto \nu^{-1/4}$, respectively.
\begin{figure*}[th]
\centering
\includegraphics[width=90mm]{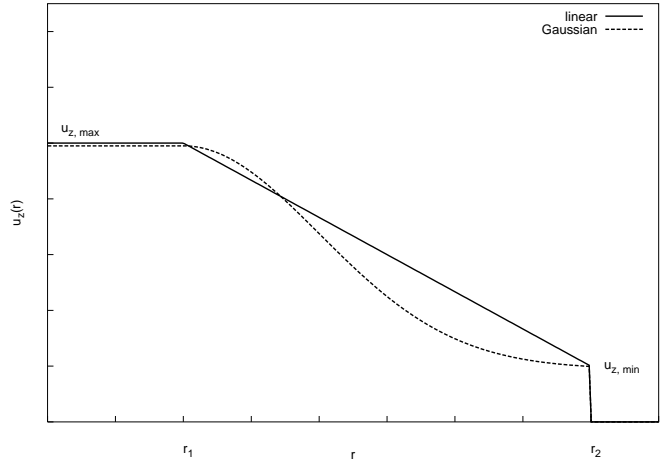}
\caption{Schematic illustration of a linearly decreasing and a Gaussian-type
velocity profile for a shear flow parallel to the jet axis, with $u_{\rm z, max}$ 
up to an inner radius $r_1$ and $u_{\rm z, min}$ at $r_2$, and $r_c = r_1$ for the
Gaussian profile.} 
\label{fig1}
\end{figure*}

\section{Acceleration timescales}
The efficiency of shear acceleration can be studied by determining the acceleration 
timescales for typical shear flow velocity profiles. In the present context we will 
concentrate on well-collimated (cylindrical) relativistic AGN jets, leaving a detailed 
analysis of expanding relativistic jet flows to the near future (Rieger \& Duffy, in 
preparation). In general, at least three different shear scenarios may be distinguished 
by means of observational and theoretical arguments \cite{rie04}: 

\subsection{Gradual shear flow along the jet axis}
As noted in the introduction, there is growing evidence today for an internal velocity 
structure parallel to the jet axis, with the simplest scenario consisting of a fast 
moving inner spine and a slower moving boundary layer. In addition, Laing et al.~(1999) 
have shown that the intensity and polarization systematics in kpc-scale FR~I jets are 
suggestive of a radially (continuously) decreasing velocity profile $v_z(r)$.
In order to estimate the particle acceleration efficiency in a longitudinal gradual 
shear flow, the following applications seem thus to be particularly interesting 
(see also Fig.~\ref{fig1}): (i) a shear flow with velocity profile decreasing linearly 
from relativistic speed $u_{\rm z, max}$ to $u_{\rm z, min}$ over a characteristic 
length scale $\Delta r = (r_2-r_1)$, and (ii) a Gaussian-type velocity profile with 
core radius $r_c$.

\subsubsection{Linearly decreasing velocity profile}
\begin{figure*}[ht]
\centering
\includegraphics[width=80mm]{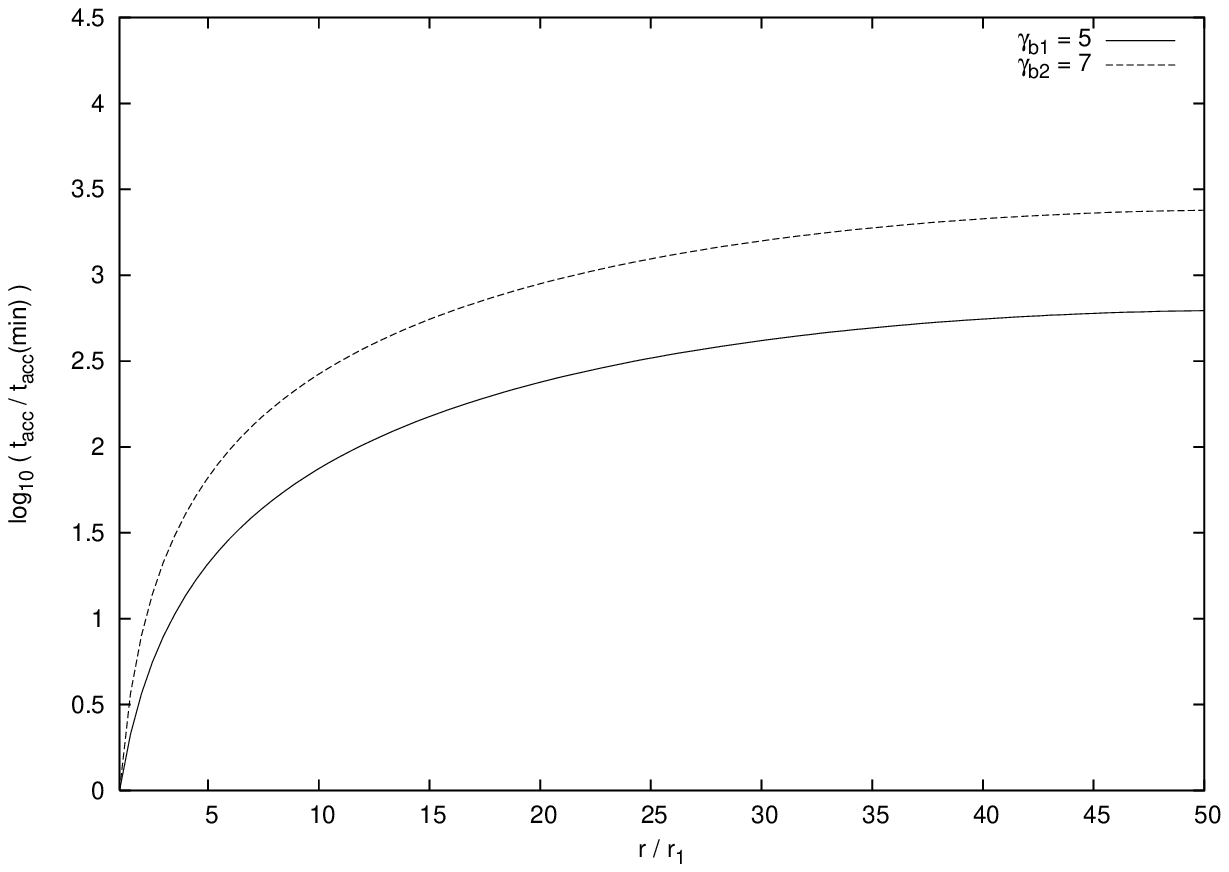}
\includegraphics[width=80mm]{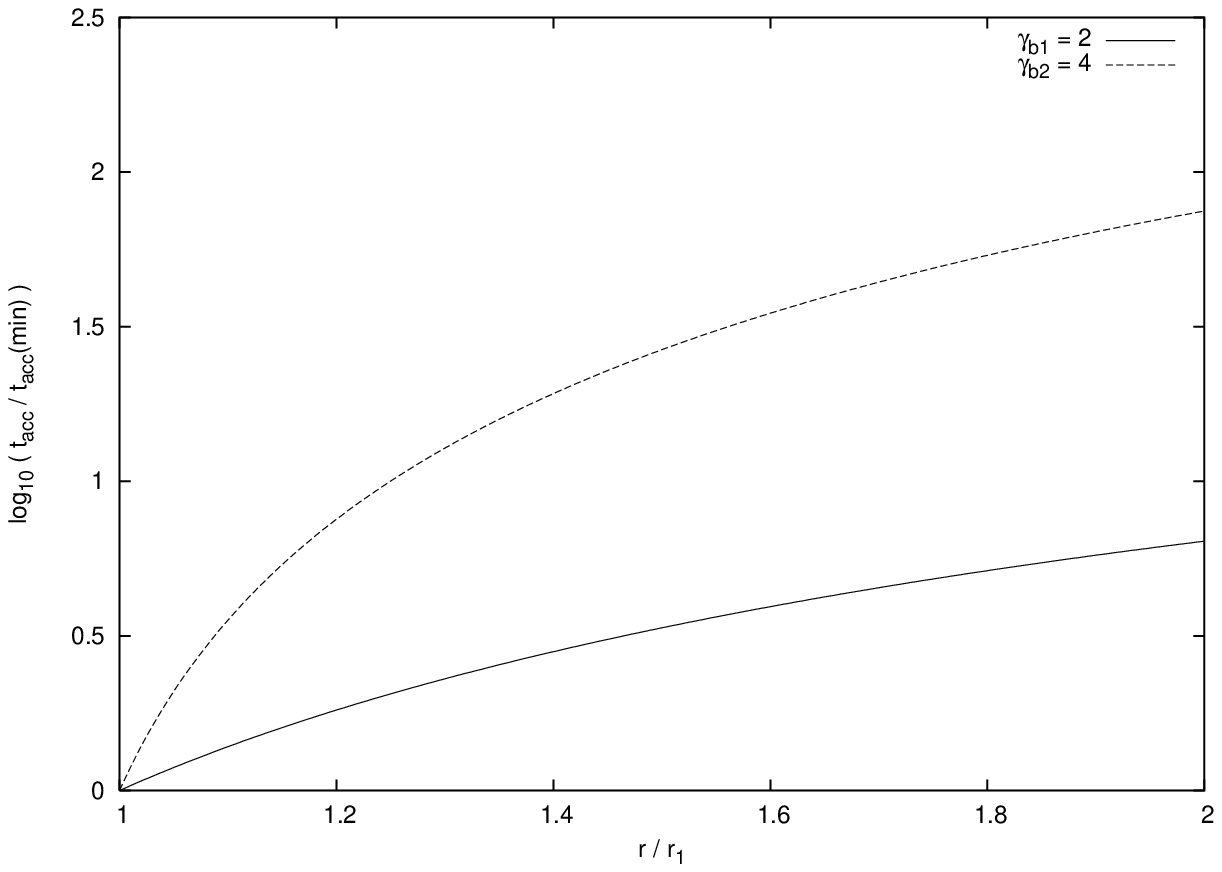}
\caption{The acceleration timescales for a gradual, parallel shear flow with 
linear decreasing velocity profile as a function of the radial coordinate.
{\it Left:} Shear flow with bulk Lorentz factors $\gamma_b(r_1)=5$ and $7$, 
respectively, $r_2=50\,r_1$ and $u_z(r_2) = 0.05\,u_z(r_1)$. The mini\-mum 
acceleration timescale is defined by $t_{\rm acc}({\rm min}) = t_{\rm acc}(r_1)$
{\it Right:} Shear flow with conditions more appropriate for FR~I jets, i.e.,
bulk Lorentz factors $\gamma_b(r_1)=2$ and $4$, respectively, $r_2=2\,r_1$ 
and $u_z(r_2) = 0.7\,u_z(r_1)$.}\label{fig2}
\end{figure*}

It can be shown (see Rieger \& Duffy~2004) that for a linearly decreasing velocity 
profile the shear acceleration timescale becomes
\beqn
 t_{\rm acc} (r) \simeq \frac{3}{\lambda}\, \frac{(\Delta r)^2}{\gamma_b(r)^2\,
             \left[1+\gamma_b(r)^2\,u_z(r)^2/c^2\right]}\, \frac{c}{(\Delta u_z)^2}
              \,\nonumber \\  
\eeqn where $\lambda(\gamma')$ is the mean free path for a particle with commoving
Lorentz factor $\gamma'$, $\Delta u_z \equiv (u_{\rm z, max} - u_{\rm z, min})$ and 
$\gamma_b(r) = 1/(1-u_z(r)^2/c^2)^{1/2}$ is the local bulk flow Lorentz factor. For 
nonrelativistic $u_{\rm z, min}$ and $\gamma_b(r_1)$ larger than a few, the minimum 
acceleration timescales is thus of order
\beq
 t_{\rm acc}(r_1) \simeq \frac{3\,(\Delta r)^2}{\gamma_b(r_1)^4\,\lambda \, c}\,.
\eeq Note, however, that if $u_{\rm z,min}$ is still relativistic with $u_{\rm z, 
min} \sim 0.7\,u_{\rm z, max}$ as suggested, for example, in the case of FR~I jets 
\cite{lai02}, the minimum acceleration timescale might easily be an order of 
magnitude larger. The evolution of the acceleration timescale as a function of $r$ 
is shown in Fig.~\ref{fig2}. The acceleration timescale increases significantly (up 
to a factor $\sim 2 \cdot 10^3$ for the application in Fig.~\ref{fig2}a) while going 
outwards with $r$ due to the decrease in flow velocity, suggesting that the higher 
energy emission will be located closer to $r_1$.\\
Any particle acceleration process will have to compete at least with radiative 
energy losses. If one requires efficient shear acceleration, this thus translates  
into an upper limit for the allowed width $\Delta r$ of the transition layer.
In the case of synchrotron radiation, for example, the cooling timescale is given
by $t_{\rm cool} \propto m^3 \,\gamma'^{-1}\,B^{-2}$, with $m$ the rest mass of the 
particle. For a gyro-dependent particle mean free path, i.e., $\lambda \sim 
r_g(\gamma')$, the condition $t_{\rm acc}(r_1) < t_{\rm cool}$ thus implies an
upper limit 
\beq 
 \Delta r \lppr \,(c^4/e^{5/2})\,\gamma_b(r_1)^2\,m^2\,B^{-3/2}\,.
\eeq 
For typical pc-scale parameters, i.e., magnetic field strength of $B \sim 0.01$ 
Gauss and $\gamma_b(r_1) \sim 10$, one thus requires a width $\Delta r_e < 0.003$ 
pc and $\Delta r_p < (m_p/m_e)^2 \,\Delta r_e$ (improving the estimate in 
\cite{rie04}) for efficient electron and proton acceleration, respectively. While 
efficient acceleration of protons is thus very likely, efficient acceleration of 
electrons on the pc-scale and below appears quite restricted, i.e., only possible if 
the transition layer is relatively thin. Note, however, that for powerful large-scale 
relativistic jets (such as the one in 3C~273) with typical parameters $\gamma_b(r_1) 
\sim 5$ and $B \sim 10^{-5}$ G, only $\Delta r_e < 30 $ pc would be required,
suggesting that efficient electron acceleration may be more likelier in large-scale 
relativistic jets. Note also, that for $\lambda$ scaling with the gyro-radius, 
$t_{\rm acc} \propto t_{\rm cool} \propto 1/\gamma'$, so that radiation losses 
are no longer able to stop the acceleration process once it has started to 
operate efficiently.

\subsubsection{Gaussian velocity profile}
Consider for comparison a Gaussian-type velocity profile, which for $r \geq r_1$
decreases as
\beq
 u_z(r) = u_{\rm z, min} + (\Delta u_z)\,\exp\left[-(r-r_1)^2 /(2\,r_c^2)\right]\,,
\eeq i.e., asymptotically approaches $u_{\rm z, min}$, where $r_c$ is the core 
radius and $\Delta u_z$ the velocity difference as defined above. The acceleration 
timescale then becomes, cf. Rieger \& Duffy~(2004),
\beqn
 t_{\rm acc}(r) =\frac{t_0\,\exp[(r-r_1)^2/r_c^2]}{(r/r_c-r_1/r_c)^2\,\gamma_b(r)^2
               \left[1+\gamma_b(r)^2\,u_z(r)^2/c^2\right]}
                \hspace{-0.5cm} \nonumber \\
\eeqn with $t_0 = 3\,c\,r_c^2/[\lambda\,(\Delta u_z)^2]$ and $r > r_1$. The minimum 
of the acceleration timescale is then no longer located at $r_1$, but still 
close-by.
\begin{figure*}[ht]
\centering
\includegraphics[width=80mm]{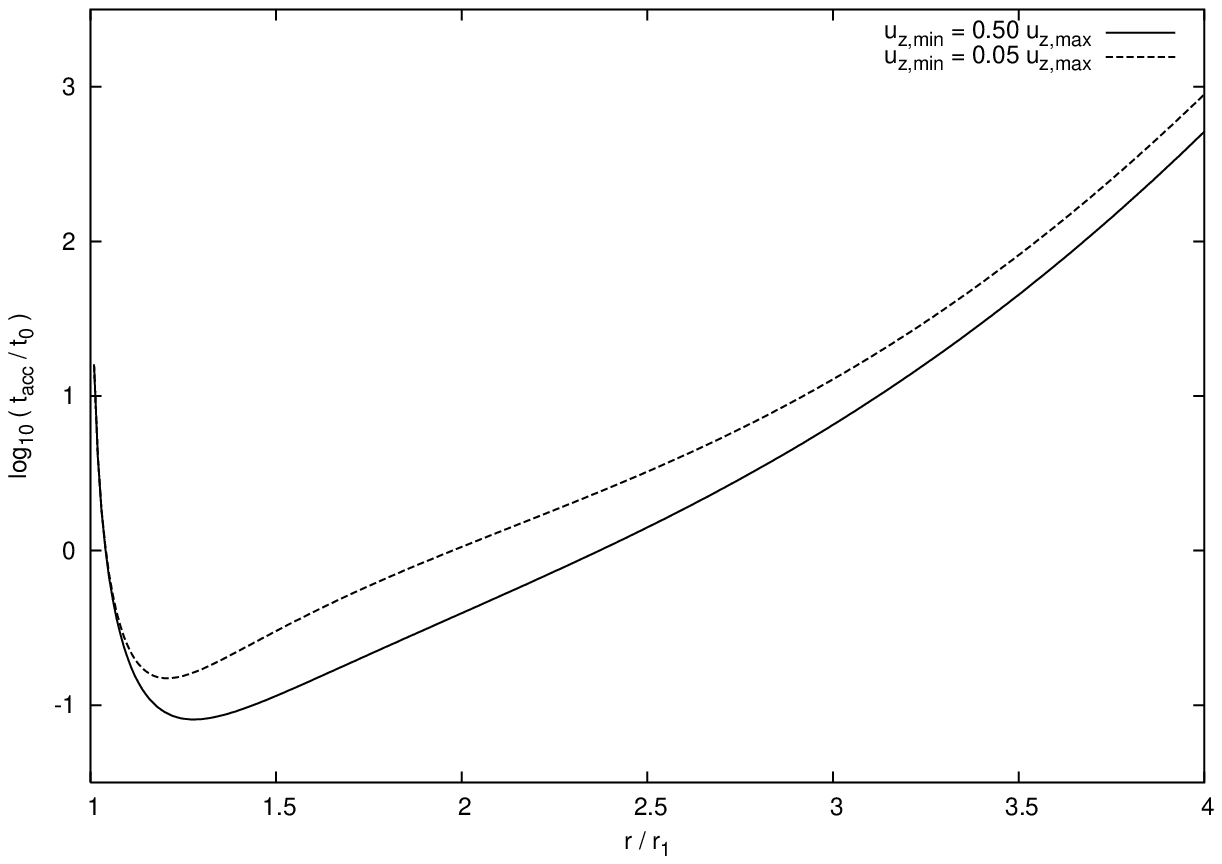}
\includegraphics[width=80mm]{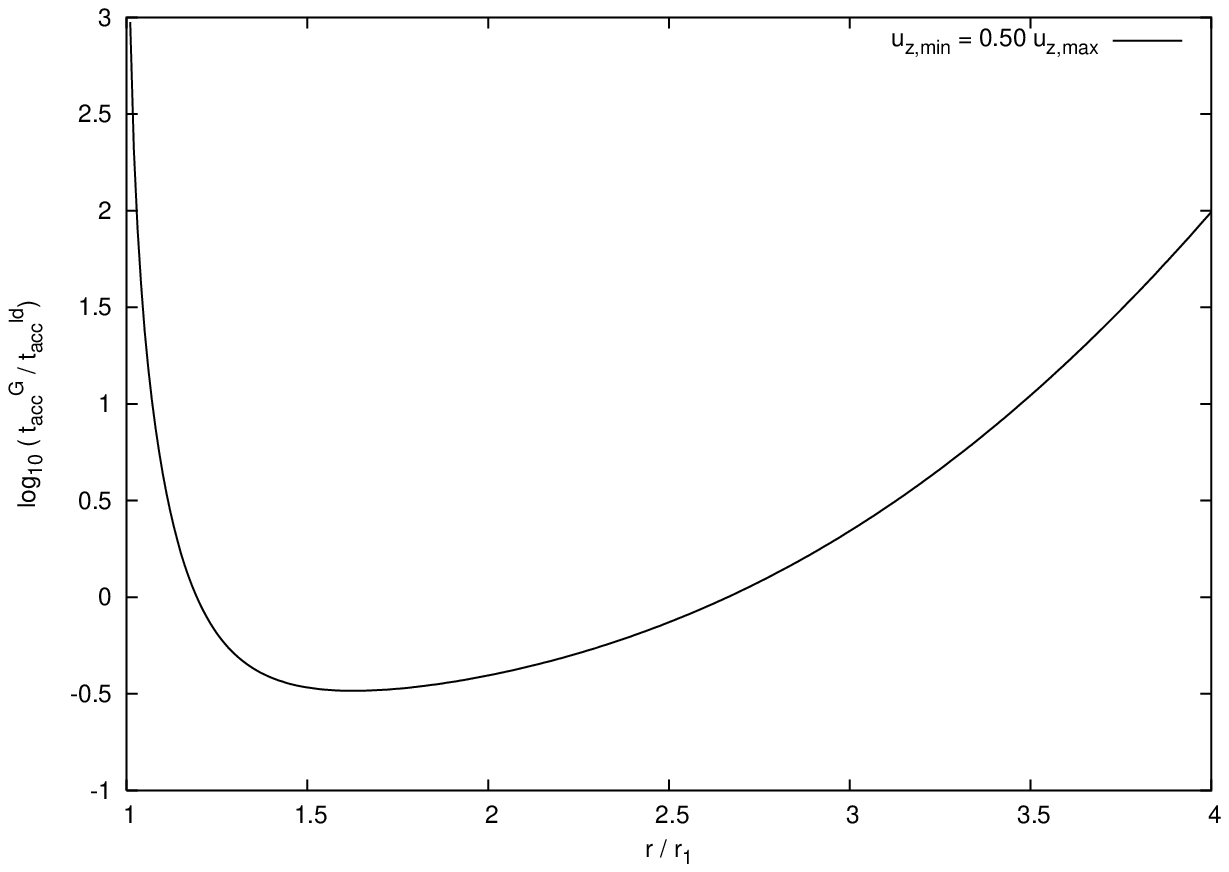}
\caption{{\it Left:} The acceleration timescales for a longitudinal gradual shear 
flow with a Gaussian-type velocity profile as a function of the radial coordinate 
for $\gamma_b(r_1) = 5$, and $u_{\rm z,min} = 0.5\,u_{\rm z,max}$ and $u_{\rm z,min} 
= 0.05\,u_{\rm z,max}$, respectively, $r_c = r_1$ and $r \leq 4\,r_1$. {\it Right:}
The ratio of the acceleration timescale $t_{\rm acc}^G$ for a Gaussian profile versus
the acceleration timescale $t_{\rm acc}^{ld}$ for a linear decreasing profile, shown
using $\gamma_b(r_1) =5$, $r_c = r_1$, $r_2 = 4\,r_1$, and $u_{\rm z,min} = 0.5\,
u_{\rm z,max}$.}\label{fig3}
\end{figure*}
The evolution of the acceleration timescale as a function of the radial coordinate 
$r$ is shown on the left hand side of Fig.~\ref{fig3}. The acceleration timescale 
now first decreases up to a minimum before eventually increasing again significantly 
due to the diminishing velocity shear. An example, comparing the acceleration timescale 
for a Gaussian with the one for a linear-decreasing velocity profile, is shown on the 
right hand side of Fig.~\ref{fig3}, illustrating the range over which a Gaussian is 
slightly preferable. Note, however, that for the chosen parameters the minimum 
Gaussian acceleration timescale is still higher by a factor of a few than the 
minimum acceleration timescale for the linear decreasing profile.

\subsection{Non-gradual shear flow along the jet axis}
Suppose a particle becomes so energetic that its mean free path $\lambda$ 
exceeds the width $\Delta r$ of the velocity transition layer. The particle may 
then be regarded as essentially experiencing a non-gradual velocity shear, i.e., 
a discontinuous jump in the flow velocity. It has been pointed out by Ostrowski 
\cite{ost90,ost98} that the jet side boundary in powerful AGN jets may naturally 
represent a relativistic example where efficient non-gradual shear acceleration 
may occur. The minimum acceleration timescale (measured in the ambient medium 
rest frame) can be estimated using Monte-Carlo simulations \cite{ost90,ost98} and 
appears to be comparable to the minimum timescale of nonrelativistic shock 
acceleration in the Bohm limit ($\lambda \sim r_g$), i.e.,
\beq
   t_{\rm acc} \sim 10\, \frac{r_g}{c} \quad\quad {\rm provided}\quad r_g > 
            \Delta r\,,
\eeq where $r_g$ denotes the particle gyro-radius. Balancing the acceleration 
timescale with the synchrotron cooling timescale then translates into an upper 
limit for the maximum attainable particle Lorentz factor of $\gamma < 4 \cdot 
10^7 \,(m/m_e)\,B^{-1/2}$. Efficient acceleration thus requires 
\beq
\Delta r < 2 \cdot 10^{-8}\,\left(\frac{m}{m_e}\right)^2\, B^{-3/2} 
           \quad {\rm  pc}\,.
\eeq 
Hence, for a pc-scale magnetic field strength of $B \sim 0.01$ G, a width of 
$\Delta r_e < 2 \cdot 10^{-5}$ pc and $\Delta r_p < (m_p/m_e)^2 \,\Delta r_e$ 
would be needed in order to allow for electron and proton acceleration, 
respectively. Efficient electron acceleration on the pc-scale thus appears  
unlikely, while efficient acceleration of protons could be well possible as 
long as the particle mean free path remains smaller than the width of the 
jet. Note that for the large-scale case with $B \sim 10^{-5}$ G one would 
still need $\Delta r_e \lppr 0.6 $ pc, suggesting that electrons are also
not accelerated efficiently on kpc-scales.

\subsection{Gradual transversal shear flow} 
While the formation of relativistic jets generally seems to be connected to the 
presence of an accretion disk, a central black hole is not necessarily required 
\cite{fen04}. If jets indeed originate as collimated disk winds, however, a 
significant amount of rotational energy of the disk may be channeled into the 
jet leading to a flow velocity field which is characterized by an additional 
rotational component (cf. \cite{rie02,rie04} for more details). In order to 
evaluate the acceleration potential associated with such a shear flow, we have 
recently analyzed the transport of energetic particles in rotating relativistic 
jets \cite{rie02}. Based on an analytical (mixed-frame) approach, and starting 
with the relativistic Boltzmann equation and a (BKG) relaxation scattering term, 
we have studied the acceleration of particles in a cylindrical jet model with 
relativistic outflow velocity $u_z$ and different azimuthal rotation profiles. 
It turned out, for example, that in the case of a simple (non-relativistic) 
azimuthal Keplerian profile, where shear effects dominate over centrifugal
effects, local power law distributions $f(p) \propto p^{-(3+\alpha)}$ are 
generated for $\tau \propto p^{\alpha}$, $\alpha >0$, whereas for more 
complex rotational profiles (e.g., flat rotation, where centrifugal effects 
become relevant) steeper spectra may be produced.\\
In order to gain insights into the acceleration efficiency, we may consider 
a flow velocity field with constant relativistic $u_z$ and azimuthal Keplerian 
rotation profile of the form $\Omega(r) = \Omega_k\,(r_{\rm in}/r)^{3/2}$, 
where $\Omega_k$ is a constant. The acceleration timescale is then of
order (see \cite{rie04})
\beq\label{eq1}
  t_{\rm acc}(r) \simeq \frac{c}{\lambda}\,\frac{1}{\gamma_b(r)^4\,(1-u_z^2/c^2)
                       \,\Omega_k^2}\,\left(\frac{r}{r_{\rm in}}\right)^3\,,
\eeq assuming the flow to be radially confined to a region $r_{\rm in} \leq r 
\leq r_j$, where $r_j$ is the jet radius and $\gamma_b(r) = 1/[1-\Omega(r)^2
\,r^2/c^2 -u_z^2/c^2]^{1/2}$ is the local flow Lorentz factor. According to 
Eq.~(\ref{eq1}) efficient particle acceleration generally requires a region 
with significant rotation. As the acceleration timescale increases considerably 
with $r$, the higher energy emission will generally be concentrated closer to 
the axis (i.e., towards smaller radii). A comparison of acceleration and cooling 
timescales indicates that efficient electron acceleration on the pc-scale is 
only possible close to $r_{\rm in}$, whereas proton acceleration is not subject 
to such a restriction. Note again that for $\lambda \propto \gamma$, the ratio 
of acceleration timescale to cooling timescale is independent of $\gamma$, 
suggesting that losses are no longer able to stop the acceleration process 
once it has started to operate efficiently.

\section{Conclusions}
There is growing evidence today that astrophysical jets may indeed be characterized 
by a significant shear within their flow velocity field. Internal jet rotation, for 
example, is likely to be present at least in the initial parts of the flow. On the
other hand, a significant and continual velocity shear parallel to the jet axis,  
is expected for most powerful jet sources. Applying the results derived above thus
suggests the following conclusions: 
\begin{itemize}
\item Gradual shear acceleration of electron occurring in small-scale relativistic 
 jets may naturally account for a steady second population of synchrotron-emitting 
 particles, contributing to the observed emission in addition to shock-accelerated 
 \cite{kirk99} ones. Depending on the level of turbulence (i.e., the momentum index
 $\alpha$ for the mean scattering time) shear-accelerated electrons may lead to
 very flat synchrotron emission spectra. On the other hand, if the Alfven velocity 
 is very high, the radio and optical emission properties of large-scale relativistic 
 jets, such as the one in 3C~273 \cite{jes01,jes05}, may be dominated by synchrotron 
 emission from relativistic electrons accelerated via second-order Fermi processes 
 (cf. \cite{sta02}). The concrete details, however, will depend on the energy scale 
 at which shear acceleration takes over \cite{rie04}.
\item Gradual and non-gradual shear acceleration processes usually work very well 
 for protons, which indicates that it may be possible to accelerate cosmic-rays to 
 ultra-high energies along powerful relativistic jets, cf. also \cite{ost98,ost00}. 
\item Shear acceleration may generally represent an efficient way of dissipating a 
 considerable part of the kinetic energy of the jet, thus possibly accounting for 
 the substantial deceleration of jets in lower-power radio sources (e.g., \cite{wang04}).
\item If a jet and its shear layer decelerates and expands significantly with 
 distance as expected, for example, in FR~I sources, the minimum shear acceleration 
 timescale will increase as well, suggesting the existence of a distance scale 
 above which shear acceleration of electrons may no longer overcome the energy losses.
\item Compared with a simple homogeneous jet model, the emission properties of 
 astrophysical shear flows may be much more complex, e.g., the inverse Compton 
 emission can be strongly boosted as each jet part will see an enhanced radiation 
 field from the other parts, cf. \cite{ghis05}.
\end{itemize}

\bigskip 
\begin{acknowledgments}
 The work of FMR is supported by a Marie-Curie Individual Fellowship
 (MCIF - 2002 - 00842). Discussions with John Kirk, Lukasz Stawarz
 and Robert Laing are gratefully acknowledged.
\end{acknowledgments}

\end{document}